\documentclass[final]{elsarticle}

\usepackage[cmex10]{amsmath}
\usepackage{url}
\usepackage[caption=false,font=footnotesize]{subfig}
\usepackage{multirow}
\usepackage[table]{xcolor}
\usepackage{enumitem}
\usepackage{hhline}
\usepackage{colortbl}
\usepackage{algpseudocode, algorithmicx}

\newcommand{\tabincell}[2]{\begin{tabular}{@{}#1@{}}#2\end{tabular}}
\newlength{\Oldarrayrulewidth}
\newcommand{\Cline}[2]{%
	\noalign{\global\setlength{\Oldarrayrulewidth}{\arrayrulewidth}}%
	\noalign{\global\setlength{\arrayrulewidth}{#1}}\cline{#2}%
	\noalign{\global\setlength{\arrayrulewidth}{\Oldarrayrulewidth}}}

\usepackage[margin=1.2in]{geometry}

\journal{}

\begin{document}

\begin{frontmatter}

\title{On Intra Prediction for Screen Content Video Coding\tnoteref{mytitlenote}}
\tnotetext[mytitlenote]{Part of the work appears at the IEEE International Conference on Image Processing 2014, with a top 10\% paper award.}

\author[uwaddress]{Haoming~Chen\fnref{myfootnote}\corref{mycorrespondingauthor}}
\fntext[myfootnote]{The work was performed when H. Chen was an intern at Samsung.}
\cortext[mycorrespondingauthor]{Corresponding author}
\ead{eehmchen@uw.edu}

\author[samsungaddress]{Ankur~Saxena}
\ead{a.saxena1@samsung.com}

\author[samsungaddress]{Felix~Fernandes}
\ead{felix.f@samsung.com}

\address[uwaddress]{Department of Electrical Engineering, University of Washington, Seattle, WA 98105 USA }
\address[samsungaddress]{Samsung Research America, 1301 East Lookout Drive, Richardson, TX 75082 USA}

\begin{abstract}
Screen content coding (SCC) is becoming increasingly important in various applications, such as desktop sharing, video conferencing, and remote education. When compared to natural camera-captured content, screen content has different characteristics, in particular sharper edges. In this paper, we propose a novel intra prediction scheme for screen content video. In the proposed scheme, bilinear interpolation in angular intra prediction in HEVC is \emph{selectively} replaced by nearest-neighbor intra prediction to preserve the sharp edges in screen content video. We present three different variants of the proposed nearest neighbor prediction algorithm: two implicit methods where both the encoder, and the decoder derive whether to perform nearest neighbor prediction or not based on either (a) the sum of the absolute difference, or (b) the difference between the boundary pixels from which prediction is performed; and another variant where Rate-Distortion-Optimization (RDO) search is performed at the encoder to decide whether or not to use the nearest neighbor interpolation, and explicitly signaled to the decoder. We also discuss the various underlying trade-offs in terms of the complexity of the three variants. All the three proposed variants provide significant gains over HEVC, and simulation results show that average gains of 3.3\% BD-bitrate in Intra-frame coding are achieved by the RDO variant for screen content video. To the best of our knowledge, this is the first paper that 1) points out current HEVC intra prediction scheme with bilinear interpolation does not work efficiently for screen content video and 2) uses different filters adaptively in the HEVC intra prediction interpolation.
\end{abstract}


\end{frontmatter}


\section{Introduction}
\label{sec:intro}
\begin{figure}[!t]
	\centerline{\subfloat[]{\includegraphics[width=0.15\textwidth]{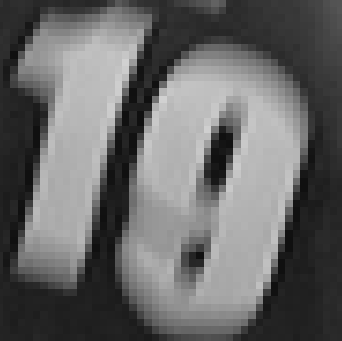}%
			\label{fig:natural}}
		\hfil
		\subfloat[]{\includegraphics[width=0.15\textwidth]{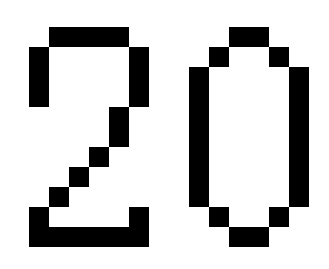}%
			\label{fig:screen}}}
	\caption{(a) Natural camera-captured image (from sequence ``BasketballDrillText'') and (b) Screen content image (from sequence ``sc\_SlideShow'')}
	\label{fig:content}
\end{figure}

Screen content coding (SCC) is widely used for various applications, such as desktop sharing, video conferencing, and remote education. The latest video coding standard H.265/HEVC provides significant compression gains over its predecessor H.264/AVC \cite{ohm2012comparison} and a lot of work are done on to improve its coding efficiency and reduce the coding complexity \cite{shen2013effective} \cite{duan2014novel} , its coding performance can be further improved for SCC as the original version 1 of HEVC video codec, or version 2 (HEVC Range Extensions) which are primarily designed for natural camera-captured content video. Currently, there is a Screen Content Coding extension work under progress in JCTVC standardization as an HEVC extension. In general, screen content video has different characteristics as compared to natural camera-captured video content, and various compression tools have been proposed for improving the coding efficiency for screen content. For example, screen content has less color intensities, and has fewer sharp edges. A palette mode \cite{lan2011intra} was proposed to represent the pixels in screen content with fewer values, so that fewer bits are needed to encode each coding block. A transform skip scheme \cite{lan2012intra} was also proposed to improve the coding performance for such video sequences. Since the DCT or DST transforms \cite{Han_ICASSP10, SaxenaICIP2011} in HEVC are, in general, not suitable for screen content video, especially at lower block sizes, transforms along horizontal and vertical directions can be skipped based on the Rate-Distortion performance. Sample adaptive prediction (SAP) or residual differential PCM (RDPCM) was proposed for horizontal and vertical intra prediction modes \cite{zhou2011ahg22, joshi2013residual}, and diagonal intra prediction modes \cite{chen2013nonrce2, saxena2013nonrce2} to improve the prediction accuracy. In this scheme, reconstructed pixels within current coding block are used to predict the current pixel. In addition, screen content also has some repeated patterns, e.g., text letters, and numbers in an article, for example in a Word or Powerpoint file. Intra block copy framework \cite{joshi2013ahg8} was proposed to utilize this spatial redundancy to further improve the compression efficiency of SCC. A multi-stage directional and temporal scheme is proposed in \cite{zhu2013screen}, in this method the intra-coding block and motion compensated block are represented as an index map. The transform skip, SAP/RDPCM and intra block copy have been adopted in the current ongoing range extension standardization for H.265/HEVC \cite{flynn2013common}.
s
In this paper, we focus on another characteristic of the screen content: sharp edges. Due to the nature of the image sensor and the effect of the camera lens, camera captured content has a smoothed boundary. However, screen content has sharp edges. A comparison of natural and screen content is shown in Fig. \ref{fig:content}. In current angular intra prediction scheme of HEVC, a bilinear filter is used to smooth the reference samples on the boundary which are used as predictors for the current block which is being encoded, to achieve a better ``smoothed'' prediction. However, for screen content video or computer-generated graphics, the content has a lot of sharp edges. In such a case, smoothing the reference pixels actually makes the prediction inaccurate as compared to the intensity values of the pixels being predicted in the current block. Hence, we propose a nearest neighbor prediction scheme in intra coding, which can derive better predictors, especially when sharp edges exist in the image. Part of the work appeared at \cite{ChenICIP2014}. In this journal paper, we provide a more comprehensive treatment of the topic, and provide various techniques for nearest neighbor prediction. We also provide detailed results, and complexity analysis of the presented coding schemes. To the best of our knowledge, this is the first paper that 1) points out current HEVC intra prediction scheme with bilinear interpolation does not work efficiently for screen content video and 2) uses different filters adaptively in the HEVC intra prediction interpolation.

The rest of the paper is organized as follows: The proposed nearest neighbor intra prediction scheme is presented in Section~\ref{sec:NNIP}. The sum of absolute difference selection method for nearest neighbor prediction is presented in Section~\ref{sec:SAD}. The pixel difference based selection method for nearest neighbor prediction is presented in Section~\ref{sec:Diff}. Section~\ref{sec:RDO} presents a rate-distortion-optimization (RDO) variant for the nearest neighborhood intra prediction. Experimental results are presented in Section~\ref{sec:Experiment} and the complexity is discussed in Section~\ref{sec:complexity}, followed finally by conclusions in Section~\ref{sec:conclusion}.


\section{Proposed Nearest-Neighbor Intra Prediction}
\label{sec:NNIP}
\begin{figure}[!t]
	\centering
	\includegraphics[width=0.4\textwidth]{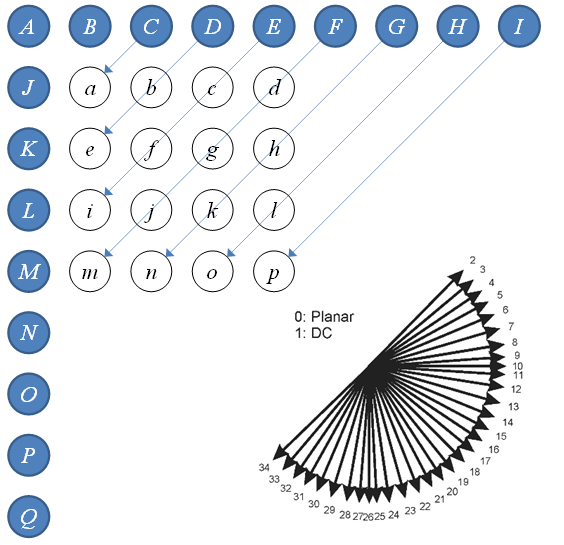}
	\caption{Right bottom figure shows 35 intra prediction modes and orientations in HEVC. Left top figure is a $4\times4$ block example of the intra prediction process. The coding block has pixels a$\sim$p. Pixels A$\sim$Q are neighboring reconstructed pixels, which are used as the reference pixels. If mode 34 (diagonal down left) is selected, pixels C$\sim$I are used to predict this block.}
	\label{fig:IntraPred}
\end{figure}

\begin{figure}[!t]
	\centerline{\subfloat[]{\includegraphics[width=0.2\textwidth]{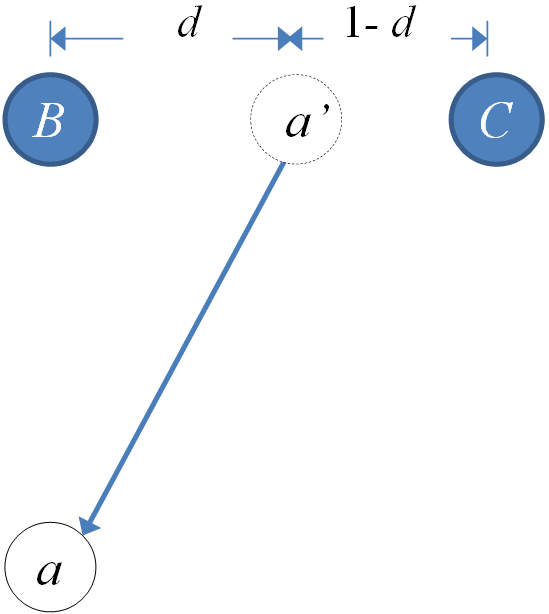}%
			\label{fig:bilinear}}
		\hfil
		\subfloat[]{\includegraphics[width=0.2\textwidth]{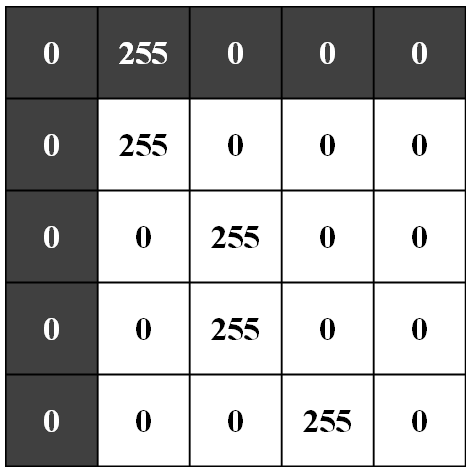}%
			\label{fig:toyEx}}}
	\caption{(a) Bilinear interpolation in HEVC intra prediction. The pixel $a'$ is the hypothetical predictor position of pixel $a$. (b) A synthetic image patch with neighboring reference samples (shaded).}
\end{figure}

In HEVC \cite{HEVCSullivan}, up to 35 intra prediction modes are supported, including Planar, DC and 33 angular modes, as shown in Fig.~\ref{fig:IntraPred}. Among the various directional modes, Horizontal, Vertical and three diagonal modes (diagonal down right, diagonal down right and diagonal up right) derive the predictors directly from the integer position reference samples along the prediction direction. For example, prediction for diagonal down left direction (mode 34) for a $4\times4$ block is shown in Fig.~\ref{fig:IntraPred}. For the oblique non-diagonal direction (modes 3$\sim$9, 11$\sim$17, 19$\sim$25 and 27$\sim$33), integer position reference samples are not available. In this case, bilinear interpolation is applied on the neighboring two samples to obtain the hypothetical prediction position, as shown in Fig.~\ref{fig:bilinear}. So the predictor of pixel $a$ is calculated as follows:
\begin{equation}
a' = (1-d)\times B + d\times C; 0 < d < 1.
\label{eq:bilinear}
\end{equation}

As described earlier in Section~\ref{sec:intro}, for screen content video, the difference in pixel values across an object boundary could be large. And also, the neighboring pixel values on the side of the edge have a very high correlation (very likely to be the same) for screen content video. In such a case, the bilinear interpolated value $a'$ may not be close to either pixels $B$ and $C$ . Instead, the non-interpolated values can provide an accurate prediction closer (even exactly the same) to the actual value of the pixel $a$. Consequently, the energy in the residual (or absolute value of the residual $a-a'$ ) will be smaller. Hence, nearest-neighbor prediction is an accurate predictor, i.e., the reference sample that is nearest to the hypothetical predictor location $a'$  can be used as the predictor, instead of the interpolated value. The predictor is derived by the following equation:
\begin{equation}
a' = B \text{ if }d < \frac{1}{2}, \text{otherwise }a' = C.
\label{eq:NN}
\end{equation}

To show the effectiveness of the nearest neighbor interpolation scheme, we show a toy example in Fig.~\ref{fig:toyEx} where a synthetic $4\times 4$ image patch (with neighboring reference samples) is shown. The original values of the $4\times4$ block, and the reconstructed pixels on the boundary are shown. The boundary pixels are used to generate the prediction for the block by using both bilinear interpolation in HEVC, and the proposed nearest interpolation. For simplicity, only directional modes 18 to 26 (see Fig.~\ref{fig:IntraPred}) are tested, since other modes are obviously unsuitable for this orientation. Table~\ref{tab:SADComparisonOfBilinearAndNearestNeighborInterpolationOnFigRefFigToyEx} shows the sum of absolute difference (SAD) between the original block and the prediction block. The minimum SAD is achieved by the nearest neighbor interpolation for mode $22$ and can be 0, which is much smaller than the minimum SAD achieved by bilinear interpolation. Such a synthetic pattern is actually very common in the screen content video, though it is unlikely to appear in natural camera captured video content. Note that one may argue that there are a lot of edge preserving interpolation techniques in the literature, e.g., bilateral filter \cite{tomasi1998bilateral} and shock filter \cite{osher1990feature}, and using NN interpolation is sub-optimal. However, considering the hardware complexity of the aforementioned techniques, and low computational complexity of NN interpolation, it is an attractive alternative. Another consideration of using the NN interpolation is the prediction error. Although some other filters could preserve the sharp edge better than bilinear interpolation, the interpolated predictor is still slightly different from the reference on either side. Since the neighboring pixels on one side of the boundary in screen content video have very high correlation (and there is no noise), they have exactly the same value. Therefore, using the nearest neighbor filter is also preferred.

\begin{table}[!t]
\footnotesize
	\caption{Sum of Absolute Differences Comparison of Bilinear Interpolation Scheme in HEVC, and the Proposed Nearest Neighbor Interpolation Scheme on the $4\times4$ Block in Fig.~\ref{fig:toyEx}.}
	\label{tab:SADComparisonOfBilinearAndNearestNeighborInterpolationOnFigRefFigToyEx}
	\centering
	\begin{tabular}{|l|l|l|} \hline
		Mode	& Bilinear &	Nearest Neighbor \\\hline
		18 &	1530	& 1530 \\\hline
		19 &	1688	& 2040 \\\hline
		20 &	1306	& 1530 \\\hline
		21 &	668	 & 1020 \\\hline
		22 &	\textbf{608 (best)}	& \textbf{0 (best)} \\\hline
		23 &	894	& 510 \\\hline
		24 &	1210	& 1530 \\\hline
		25 &	1402	& 1530 \\\hline
		26 &	1530	& 1530 \\\hline
	\end{tabular}
\end{table}

To further evaluate the performance of the proposed nearest-neighbor interpolation on real video sequences, we run a simulation as follows: we simply replace the bilinear interpolation with the proposed Eqn.~\eqref{eq:NN} for all blocks. We encoded various video sequences as all intra frames in Table~\ref{tab:results_lossy}, since the proposed method is for intra coding. Full details about these sequences, and their GOP size, Intra period, frame rate, dimensions etc., are in \cite{flynn2013common} and \cite{SaxenaRCE3}, and described in more details in Section~\ref{sec:Experiment}. The Bj{\o}ntegaard-Delta rate (BD-rate) \cite{bjontegard2001calculation} comparison of Luma component in this case is shown in Table~\ref{tab:results_lossy} method \ref{all} ``All Blocks''. From the simulation results, the nearest neighbor interpolation saves the BD-bitrate up to 12.9\% for the sequence ``sc\_cg\_twist\_tunnel'' and 2.2\% in average for screen content sequences. However, note that there is an 10.2\% BD-bitrate increase for the sequence ``BasketballDrillText'' which is primarily natural camera captured content with a strip of text in the bottom, and significant BD-bitrate increase for almost all natural sequences, which is not desirable. The reason for such a loss is that for natural camera captured content, bilinear interpolation performs efficiently, as the pixels along some edge or object boundary are smoothed. It should be noted that for a purely smooth area, e.g., white background in sequence ``sc\_wordEditing'', the best mode is always DC or planar mode, so replacing bilinear interpolation with NN interpolation makes no difference.

In theory, one could signal a flag in the Sequence Parameter Set (SPS) or frame-header to indicate that a sequence (or frame) contains screen content, and consequently the entire sequence (or frame) should use NN intra prediction. Unfortunately, such a scheme would fail because video frames typically consist of mixed screen content and camera-captured content.

Hence, applying the nearest neighbor prediction selectively based on the content is desirable, as it works well on most screen content video. In this paper, we present the following three variants to ``selectively'' apply the proposed nearest neighbor prediction, and discuss the pros and cons of the three techniques.
\begin{enumerate}
	\item Sum of absolute difference (SAD) based selection criteria (at pixel level): As the reconstructed reference pixels are available at both the encoder and the decoder, an implicit criteria is  derived based on these reference samples to determine whether to use bilinear interpolation, or the nearest neighbor prediction scheme;
	\item Pixel difference based selection criteria (at pixel level): an implicit method similar to criteria (1), but (1) is a block-level criteria and (2) is a pixel-level;
	\item Rate-Distortion based selection criteria (at block level): A rate-distortion optimized decision is made for each coding block at the encoder, and a flag is signaled to the decoder side to select between bilinear interpolation, or the nearest neighbor prediction scheme.
\end{enumerate}

\section{Sum of Absolute Difference Based Selection Variant}
\label{sec:SAD}
In general, screen content video is not as smooth as the natural content. Hence, the sum of absolute difference (SAD) can be a good indicator showing whether the content is closer to screen content video, or natural content. For a block, SAD is computed as follows:
\begin{equation}
\text{SAD} = \sum_i{|x_i - \bar{x}|}
\label{eq:SAD}
\end{equation}
where $x_i$ is the value of reference samples (e.g., $A\sim Q$ in Fig.~\ref{fig:IntraPred}), and $\bar{x}$ is the mean of those samples. If the SAD is small, implying the variation in the current block is less, and the block is smooth, it is more likely to be natural content, and bilinear interpolation can be used. On the other side, if the SAD is large, it is more likely to have sharp edges, which implies that nearest neighbor interpolation can be used. When applying this method, we use a threshold-based selection as follows:

\begin{algorithmic}
	\If {$\text{SAD} \geq threshold$}
	\State apply nearest neighbor interpolation on all pixels within current block
	\Else
	\State retain bilinear interpolation
	\EndIf
\end{algorithmic}

In intra prediction on a $N\times N$ block, the number of reference samples could be $4\times N+1$. For each intra prediction mode, not all of the reference samples are needed. As an example, for a $4\times 4$ block, we categorize these modes into three different sets, shown in Table \ref{tab:CategorizedReferenceSamplesForDifferenceIntraPredictionModesInSADCalculation}. For different categories, we calculate the SAD of corresponding reference samples.

\begin{table}[!t]
\footnotesize
	\caption{Categorized Reference Samples for Difference Intra Prediction Modes in SAD Calculation}
	\label{tab:CategorizedReferenceSamplesForDifferenceIntraPredictionModesInSADCalculation}
	\centering
	\begin{tabular}{|l|l|l|} \hline
		Category	& Modes	& Samples (as in Fig. \ref{fig:IntraPred}) \\\hline
		1	& 3-9	& J $\sim$ Q \\\hline
		2	& 11-17, 19-25	& A $\sim$ E, J $\sim$ M \\\hline
		3	& 27-33	& B $\sim$ I \\\hline
	\end{tabular}
\end{table}

Note that a similar block-level criteria such as variance, and histogram of the pixel intensity can also be used to classify the screen content and nature content. Considering the implementation complexity in the hardware (for example, taking squares in variance computation; and sorting in histogram calculation), we use the SAD criteria in our algorithm. The best SAD threshold is needed to maximize the coding gain. Based on our simulation results, we select 64 as the threshold which is also friendly to the hardware implementation, as it is a power of 2. This threshold is for 8-bit sequences, while for 10-bit or 12-bit sequences, the threshold is normalized appropriately (256 for 10-bit video etc.). The details of the experiments are discussed in Section~\ref{sec:Experiment}.

\section{Pixel Difference Based Selection Variant}
\label{sec:Diff}
The SAD based method is a block-level variant, which means that all the pixels within a coding block will use the same interpolation method: either bilinear interpolation or nearest neighbor interpolation. In this section, we propose another implicit scheme based on the reference pixel differences that adaptively applies the nearest-neighbor interpolation or the bilinear interpolation. This method is pixel-level, so different pixels within a coding block can have different interpolation methods. Specifically, the proposed algorithm is described as follows:
\begin{algorithmic}
	\If {$|B-C| \geq threshold$}
	\If {($d < 1/2$)}
	\State $a' = B$
	\Else
	\State $a' = C$
	\EndIf
	\Else
	\State retain bilinear interpolation
	\EndIf
\end{algorithmic}
where pixels $B$, $C$, $a'$, and distance $d$ are as shown in Fig.~\ref{fig:bilinear}. In our experiments, we used the threshold as the average value of the dynamic range of the pixels. For example, for 8-bit sequences with pixel values from 0 to 255, the threshold set is 128. The rationale of this scheme is that when the reference pixel values differ significantly, they typically would belong to sharp edges/regions in screen content video, and nearest neighbor interpolation will perform better than bilinear interpolation. On the other hand, if the pixel difference is small, then the region is smooth, and is belongs to natural camera-captured video content. In such a case, we retain the bilinear interpolation technique in HEVC.

In both SAD-based method and pixel-difference based method, since the reference sample can be derived identically at the decoder side, there are no overhead bits to let the decoder know about whether to use nearest neighbor prediction or not. However, these two methods are typically not rate-distortion optimized, and the decisions of using a particular interpolation might not be the optimal in R-D sense. Another issue with the SAD-based and pixel-difference based methods is that it requires additional logic and the fast RDO based on original pixels (instead of reconstructed pixels) cannot be done. We next present the Rate-Distortion method for using nearest neighbor prediction in the next section.

\section{Rate-Distortion Optimized Variant}
\label{sec:RDO}
In the proposed Rate-Distortion based scheme, we evaluate the bilinear and nearest neighbor predictions on one block (both luma and chroma components), and then select the prediction scheme that results in less R-D cost. For different possible intra prediction modes, the same R-D search is applied, and a globally optimized combination of (intra mode, interpolation) is selected. Since the intra mode information is already available at both the encoder, and decoder, we only need to encode one extra flag for the interpolation information. In this section, we further discuss the techniques in the R-D search and entropy coding of interpolation flag.
\subsection{Techniques in R-D search}
In R-D search for the best combination, since the R-D cost for bilinear interpolation scheme has already been calculated in the encoding process, a straightforward way is to search all the 34 intra prediction modes with nearest neighbor interpolation on both luma and chroma components. However, this increases the search complexity significantly at the encoder; and encoding/signaling of the interpolation flag, which actually is not always required.  We find among all these modes, some modes can altogether be skipped for the search, and also the interpolation for chroma can be derived implicitly from the corresponding luma components. Hence, the search complexity can be reduced. Specifically, we use the following techniques as described next:
\label{sec:techinRD}
\begin{enumerate}
	\item Skip the interpolation search for the Planar, DC, Horizontal, Vertical and three diagonal modes, since the integer position prediction is already applied to these modes, and bilinear interpolation and nearest neighbor prediction scheme are identical for these modes.
	\item As the intra mode information is derived prior to interpolation information in the decoder side, no signaling bits are required for describing the interpolation scheme for the Planar, DC, Horizontal, Vertical and three diagonal modes.
	\item For intra prediction of the chroma components, there are five modes available, mode 0 to mode 3 are Planar, Vertical, Horizontal, DC and one diagonal mode (mode 34). When Mode 4, the ``derived mode'' is selected, the same mode is used as the corresponding luma component. Hence, for the signaling of interpolation information, the ``derived interpolation'' can be used. If the mode 4 is selected in chroma intra prediction, the same interpolation method is used in the chroma. In this way, no additional interpolation bits are needed for chroma components.
	\item Restricting the proposed intra prediction on only $4\times4$ blocks, as the additional advantage in terms of compression gains of applying nearest neighbor prediction on larger blocks is marginal as we will show in Section~\ref{sec:Experiment}. Note that, in general, for screen content, when there are sharp edges, most of the times $4\times4$ blocks will be chosen, and hence it is sufficient to apply nearest neighbor prediction on these smaller blocks only. Note that if an encoder does not support $4\times4$ blocks, then the prediction search will be required to be performed at the lowest available block size, e.g., $8\times8$.
	\item On the block size of $4\times4$, we further find some modes can be merged under the nearest neighbor interpolation. Mode 9 and 11 have the same predictors as mode 10 (horizontal). Mode 25 and 27 have the same predictors as mode 26 (vertical). So that the search for mode 9, 11, 25 and 27 can also be skipped.
	\item To reduce the intra mode decision complexity, some fast mode decision methods are used. For example, in HEVC, some candidate modes are selected based on the SAD cost, instead of the real Rate-Distortion cost. And then the full R-D search (based on the actual bitrate and distortion) is run on those modes, so that the searching time is reduced greatly. Our algorithm is compatible to such a scheme: the RDO-based Bilinear/NN interpolation selection is applied on these candidate modes, instead of all possible modes shown in Table \ref{tab:CategorizedReferenceSamplesForDifferenceIntraPredictionModesInSADCalculation}. This method doesn't reduce any overhead of encoding the interpolation and may result in an locally optimal solution, but it is a good trade-off between the encoding time and the performance.
\end{enumerate}

\subsection{CABAC Coding of Interpolation Modes}
After deriving the best combination of (intra mode, interpolation) leading to the minimum R-D cost, the next step is to encode the interpolation into the bitstream. We assign one flag for the interpolation (0: bilinear, 1: nearest neighbor) following the intra prediction mode and use CABAC \cite{marpe2003context} to encode this bit. For the selection of context model, we investigate two methods:
\begin{enumerate}
	\item \label{enum:onecontext} Use only one context model.
	\item \label{enum:threecontext} Use three context models based on the interpolation methods on the left and top neighboring coding blocks.
\end{enumerate}
In fact, the spatial neighboring dependent model context derivation method \ref{enum:threecontext}, is used for some flags in HEVC, e.g., skip\_flag and split\_flag. This method works when the flag is not independent to the corresponding flags in the neighboring blocks. In our scenario, one block is very likely to use the nearest neighbor interpolation when its neighboring blocks use this interpolation. Suppose the interpolation flag for up and left coding blocks are flag\_up and flag\_left (flag\_up, flag\_left $\in$ \{0, 1\}), then the context mode index for encoding current block is
\begin{equation}
\text{context\_mode\_index = flag\_up + flag\_left}
\label{eq:contextupdate}
\end{equation}
If one of up and left coding blocks is not available, the corresponding flag is set to 0 as well. In the decoder side, the same context mode index can also be derived based on the neighboring coding blocks, so the same context model is used to decode the interpolation mode. According to our simulation results, about 0.1-0.2\% additional coding gain can be achieved by using the multiple context models method as compared to a single context.

\section{Experimental Results}
\label{sec:Experiment}
We encoded full length sequences (which had 120 to 600 images, all are YUV format) and various resolutions varying from 832$\times$480 to 1920$\times$1080. Some main test settings are listed in Table~\ref{tab:TestConditions}, and full details about these sequences, and their GOP size, Intra period, frame rate, dimensions etc., are in \cite{flynn2013common} and \cite{SaxenaRCE3} . The anchor (reference) was HM-12.0+RExt-4.1 \cite{hm12}, the HEVC reference software for developing SCC, with bilinear interpolation being used as default. Note that the state-of-the-art tools, such as transform skip, sample adaptive prediction and intra block copy are already implemented in it, and also enabled in our experimental tests. The performance of the proposed nearest neighbor interpolation scheme is evaluated for the following 3 settings: All Intra (AI), Random Access (RA), and Low-Delay Bi-direction prediction (LB) settings.  In the AI setting, all the images were encoded as Intra, while RA setting had periodic Intra frames; and the LB settings had only the first frame as Intra. Both lossy and lossless coding settings are tested. These video sequences are being tested as part of HEVC standardization, including both natural content video and screen content video sequences. Note that we present, here the results for only the evaluation of the proposed interpolation scheme and retain all other test settings in \cite{flynn2013common} and \cite{SaxenaRCE3}.

\begin{table}[!t]
\footnotesize
	\caption{Test Conditions}
	\label{tab:TestConditions}
	\centering
	\begin{tabular}{|l|l|}\hline
		Software	Version & HM-12.0+RExt-4.1 \\\hline
		Intra period	& \tabincell{l}{All Intra (AI): 1 \\ Random Access (RA): 16, 24, 32, 64 \\ Low Delay B (LB): -} \\\hline
		QP & \tabincell{l}{22,27,32 and 37 (Lossy, Main Tier)\\ 0 (Lossless)}\\\hline
	\end{tabular}
\end{table}

\subsection{Lossy Coding Results}
\label{sec:lossy}
The results for lossy coding are provided in Table \ref{tab:results_lossy}. We present the results for following 5 different methods:
\begin{enumerate}[label={(\arabic*)}]
	\item \label{all} Nearest neighbor prediction applied on all prediction blocks from $4\times4$ to $64\times64$;
	\item \label{4x4} Nearest neighbor prediction applied only on blocks of size $4\times4$;
	\item \label{sad} The nearest neighboring filter is selected adaptively on $4\times 4$ blocks based on the SAD criteria (threshold = 64);
	\item \label{diff} Nearest neighbor prediction is selected adaptively on $4\times 4$ blocks based on the difference of the neighboring references (threshold = 128);
	\item \label{rdo} Nearest neighbor prediction is selected based on the R-D Optimization search on the encoder side; one extra flag is included in the bit stream and counted into the total bits.
\end{enumerate}

Note that some test sequences are 10-bit sequences, and to compare with the thresholds we used values of 64 * 4 = 256 and 128 * 4 = 512 in methods \ref{sad} and \ref{diff} to appropriately normalize to the dynamic range of 10-bit video.

\begin{table*}[!t]
\tiny
	\caption{BD-Bitrate Savings for Luma Component (in \%'s) When the Proposed Nearest Neighbor Intra Prediction Applied under Difference Schemes. Note That Negative BD-Rate Means Compression Gain. (Shaded Number $\geq$ 1.0)}%
	\label{tab:results_lossy}%
	\centering
	\begin{tabular}{|c|c|c|c|c|c|c|c|c|c|c|c|c|} \hline
		\multirow{2}{*}{Category} & \multirow{2}{*}{Sequences}
		& \tabincell{c}{\ref{all} All \\blocks}
		& \tabincell{c}{\ref{4x4} Only \\ 4$\times$4  \\ blocks}
		& \multicolumn{3}{c|}{\tabincell{c}{\ref{sad} SAD variant\\ on only 4$\times$4 \\ blocks}}
		& \multicolumn{3}{c|}{\tabincell{c}{\ref{diff} Pixel difference \\ variant on \\ only 4$\times$4 blocks}}
		& \multicolumn{3}{c|}{\tabincell{c}{\ref{rdo} R-D based \\ variant on \\ only 4$\times$4 blocks}} \\\hhline{|~|~|-|-|-|-|-|-|-|-|-|-|-|}
		&       						& AI    									& AI    										& AI    	& RA    	& LB	& AI    	& RA    	& LB    	& AI    	& RA    	& LB		\\\hhline{|=============|}
		\multirow{11}{*}{\tabincell{c}{Screen \\ Content \\ (All 4:4:4 \\ format)}}
		& ChinaSpeed 				& 0.3  										& -0.7   					&-0.6&-0.3&-0.1					& -1.1    &  -0.6   & -0.3  	& -2.1  	& -1.2 	  & -0.8 	\\\hhline{|~|-|-|-|-|-|-|-|-|-|-|-|-|}
		& SlideEditing 			& -1.0  									& -0.8  					&-0.2&-0.3&0.1					& -0.3  	&  -0.3   &  0.2  	& -1.1  	& -1.1    & -0.7  \\\hhline{|~|-|-|-|-|-|-|-|-|-|-|-|-|}
		& sc\_cad\_waveform & -3.1  									& -2.8  					&-0.4&-1.0&-1.9					& -0.6  	&  -1.3   & -5.4    & -3.1  	& -3.6    & -7.6  \\\hhline{|~|-|-|-|-|-|-|-|-|-|-|-|-|}
		& sc\_pcb\_layout 	& -7.1 										& -6.8 						&-0.5&0.4&-0.4					& -1.0  	&  -1.2   & -0.2    & -7.4 		& -6.5 	  & -4.3  \\\hhline{|~|-|-|-|-|-|-|-|-|-|-|-|-|}
		& sc\_ppt\_doc\_xls & -2.9  									& -2.9  					&-0.9&-0.9&-0.4					& -1.4  	&  -1.6   & -0.9    & -3.4    & -3.1 	  & -2.8  \\\hhline{|~|-|-|-|-|-|-|-|-|-|-|-|-|}
		& sc\_programming 	& \cellcolor{gray!50}1.6  & -0.0  					&-0.2&-0.1&0.0					& -0.5  	&   0.1   & -0.3    & -1.2  	& -0.6 	  & -0.4  \\\hhline{|~|-|-|-|-|-|-|-|-|-|-|-|-|}
		& sc\_SlideShow 		& \cellcolor{gray!50}4.7  & \cellcolor{gray!50}2.3 &0.0&-0.1&-0.1  	&  0.1 		&  -0.1   & -0.3    & -0.6  	& -0.6 	  & -0.7  \\\hhline{|~|-|-|-|-|-|-|-|-|-|-|-|-|}
		& sc\_web\_browing 	& -0.5  									& -1.3  					&0.0&-0.1&-0.5					& -0.2  	&  -0.2   & -0.8    & -1.4  	& -1.3 		& -0.5  \\\hhline{|~|-|-|-|-|-|-|-|-|-|-|-|-|}
		& sc\_wordEditing 	& -1.1  									& -1.0  					&-0.7&-0.1&-0.1					& -0.4  	&  -0.6   & 0.5    & -1.5  	& -1.1 		& -0.5  \\\hhline{|~|-|-|-|-|-|-|-|-|-|-|-|-|}
		& sc\_twist\_tunnel & -12.9 									& -11.6 					&-1.2&-1.1&-1.1					& -5.4 		&  -5.2   & -5.0    & -11.5 	& -8.7 		& -8.6  \\\Cline{0.9pt}{2-13}
		& \textbf{Average} 	& \textbf{-2.2} 					& \textbf{-2.6} &\textbf{-0.5}&\textbf{-0.4}&\textbf{-0.5} & \textbf{-1.1} & \textbf{-1.1} & \textbf{-1.3}  	& \textbf{-3.3} & \textbf{-2.8} & \textbf{-2.7} \\\hhline{|=============|}
		\multirow{4}{*}{\tabincell{c}{Mixture of \\ Screen and \\Natural \\ Content}}
		& BasketballDrillText (4:2:0) 	& {\cellcolor{gray!50}}10.2  	& \cellcolor{gray!50}4.4  &0.0&-0.1&0.0 		&  0.3  &  0.0   &  0.0  & -0.1  &  0.0		&  0.0	\\\hhline{|~|-|-|-|-|-|-|-|-|-|-|-|-|}
		& sc\_map (4:4:4) 							& {\cellcolor{gray!50}} 3.0   & \cellcolor{gray!50}1.1  &0.0&0.1&0.0  	&  0.1  & -0.1   &  0.1  & -0.2  &  -0.3 	&  0.0 	\\\hhline{|~|-|-|-|-|-|-|-|-|-|-|-|-|}
		& sc\_vc\_doc\_sharing (4:4:4)	& -1.9  											&-1.7  										&-0.7&-0.6&-2.5		& -1.2  & -1.1   & -0.9  & -2.4  &  -2.2 	& -0.4	\\\Cline{0.9pt}{2-13}
		& \textbf{Average} 			& \textbf{3.8} 								& \textbf{1.2} 						&\textbf{-0.2}&\textbf{-0.2}&\textbf{-0.8}		& \textbf{-0.3} & \textbf{-0.4} & \textbf{-0.3} & \textbf{-0.9} & \textbf{-0.8} & \textbf{-0.2}\\\hhline{|=============|}
		\multirow{6}{*}{\tabincell{c}{Natural \\Content}}
		& Kimono (4:4:4)							& \cellcolor{gray!50}3.3   		& 0.5   							&0.0&0.0&0.0			& 0.0     	&  0.0    &	0.0		& 0.0     & 0.0		& 0.0	\\\hhline{|~|-|-|-|-|-|-|-|-|-|-|-|-|}
		& ParkScene (4:2:0)					& \cellcolor{gray!50}2.0     	& 0.7   							&0.0&0.0&0.1			& 0.0     	&  0.0    &	0.1		& 0.0     & 0.1		& 0.1	\\\hhline{|~|-|-|-|-|-|-|-|-|-|-|-|-|}
		& EBUHorse  (4:2:2)						& \cellcolor{gray!50}3.0   		& \cellcolor{gray!50}1.4 &0.1&0.1&0.1  	& 0.1   		&  0.1    & 0.1		& 0.0     & -0.1	& 0.1	\\\hhline{|~|-|-|-|-|-|-|-|-|-|-|-|-|}
		& EBUWaterRocksClose (4:2:2) 	& \cellcolor{gray!50}3.2   		& \cellcolor{gray!50}1.9 &0.4&0.5&0.5  	& 0.5     	&  0.5    & 0.5		& 0.0     &  0.0 	& 0.0	\\\hhline{|~|-|-|-|-|-|-|-|-|-|-|-|-|}
		& EBURainFruits (4:2:2)			& \cellcolor{gray!50}2.6   		& 0.8   								 &0.0&0.0&0.0		& 0.0    		&  0.0    &	0.0 	& 0.0     &  0.0	& 0.0	\\\Cline{0.9pt}{2-13}
		& \textbf{Average} 		& \textbf{2.8} 								& \textbf{1.1} 					 &\textbf{0.1}&\textbf{0.1}&\textbf{0.1}		& \textbf{0.1} & \textbf{0.1} & \textbf{0.1} & \textbf{0.0} & \textbf{0.0}	& \textbf{0.0}\\ \hhline{|=============|}
		\multirow{2}{*}{Summary} 	& Enc. Time (\%) & -    & -   & 101    &  100     & 100 & 101    &  101     & 101    &   108    & 100   & 100 	\\
		& Dec. Time (\%) & -    & -   & 100    &  100     & 100 & 102    &  101     & 101    &   100    & 98    & 99 	\\\hline
	\end{tabular}
\end{table*}%

In Table~\ref{tab:results_lossy}, the results for the five above methods are presented and the BD-rate is shown for Luma component. For our proposed methods, time ratio is used to compare the encoding and decoding time between the anchor and our proposed methods, which is calculated as follow:
\begin{equation}
\text{Encoding Time Ratio} = \frac{\text{Encoding Time of Proposed Method}}{\text{Encoding Time of Anchor}}
\label{eq:enctimeratio}
\end{equation}
\begin{equation}
\text{Decoding Time Ratio} = \frac{\text{Decoding Time of Proposed Method}}{\text{Decoding Time of Anchor}}
\label{eq:dectimeratio}
\end{equation}
From the results, we have the following remarks:
\begin{enumerate}
	\item Restricting the proposed interpolation on $4\times 4$ achieves most of the gains in screen content video and reduces the loss in natural content video. So, we apply the ``selective'' interpolation criteria for only $4\times 4$ blocks.
	\item The SAD based and threshold based pixel difference criterias works well on almost all sequences, and there are some slight losses on the sequence ``EBUWaterRocksClose'' since we have not optimized the threshold for each sequence separately.
	\item On most sequences, the threshold based criteria outperforms the SAD based method. Among all these three methods, the RDO performs best.
	\item The Rate-Distortion based scheme performs well on all sequences, and coding gains of 3.3\% (AI),  2.8\%(RA), and 2.7\% (LB) on average are obtained. Further, there is no loss on camera-captured content.
\end{enumerate}

\begin{figure*}[!t]
	\centerline{\subfloat[RD curve of ChinaSpeed]{\includegraphics[width=0.32\textwidth]{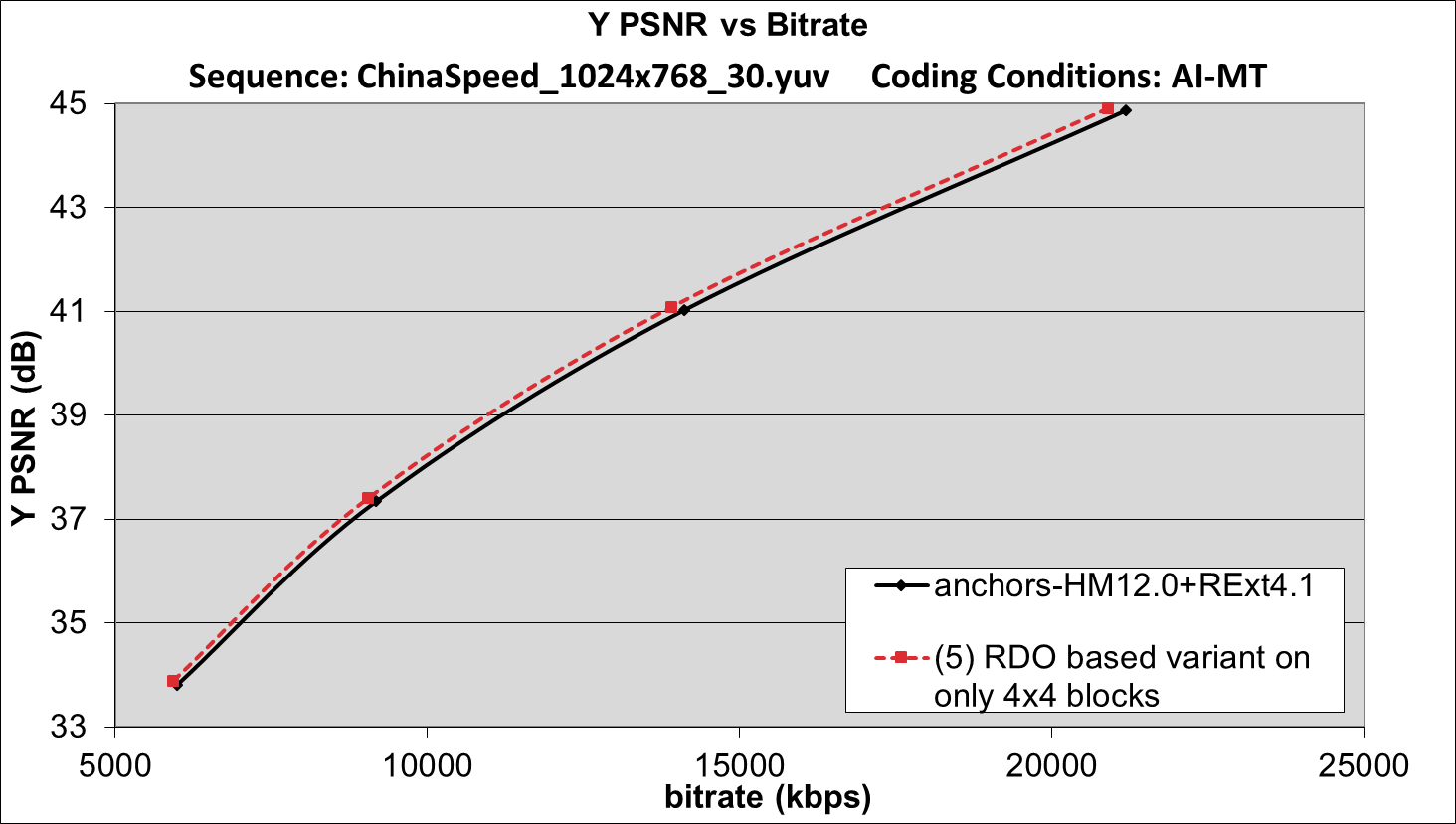}%
			\label{fig:rdc_ChinaSpeed}}
		\hfil
		\subfloat[RD curve of sc\_twist\_tunnel]{\includegraphics[width=0.32\textwidth]{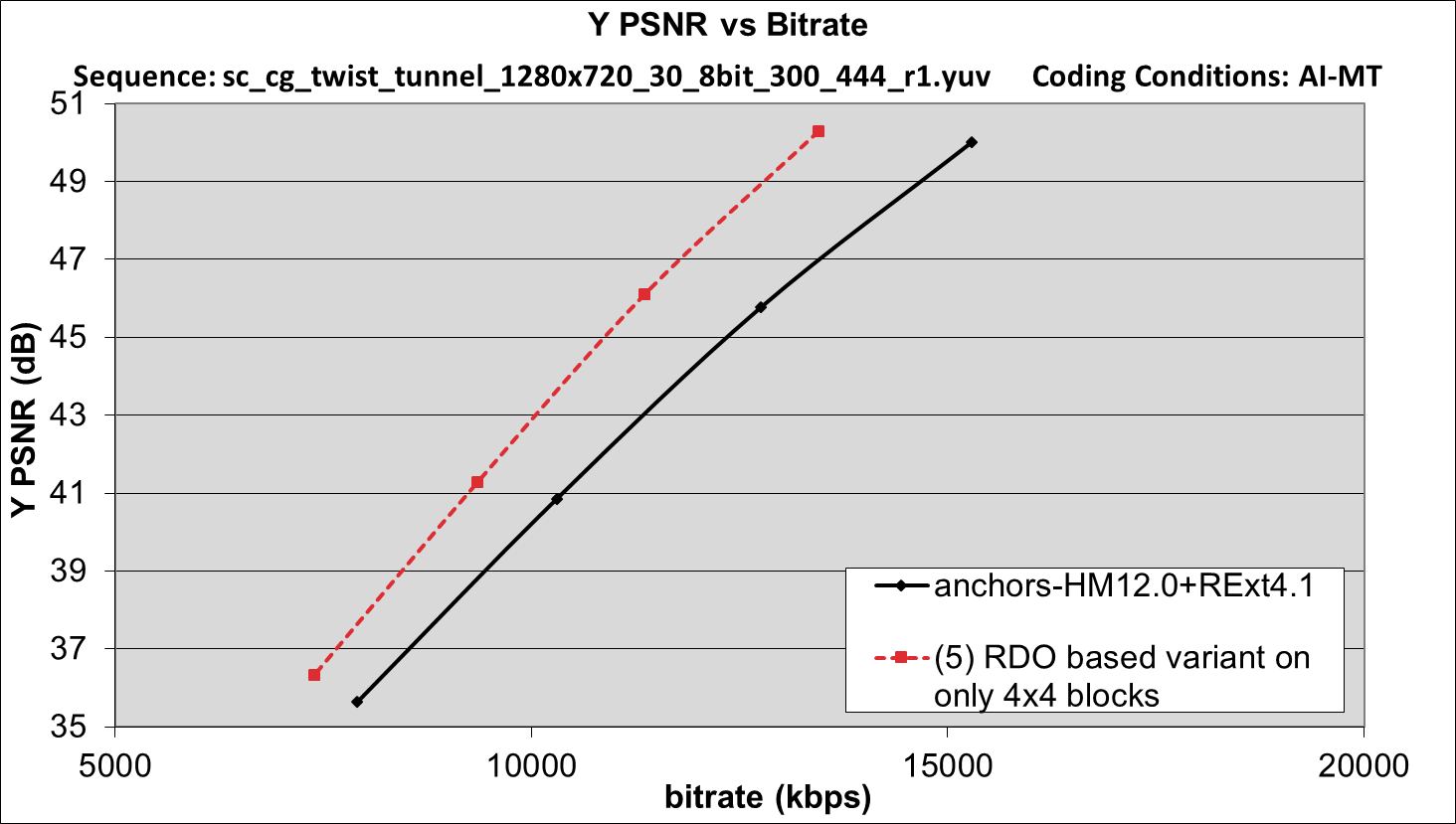}%
			\label{fig:rdc_cg}}
		\hfil
		\hfil
		\subfloat[RD curve of sc\_vc\_doc\_sharing]{\includegraphics[width=0.32\textwidth]{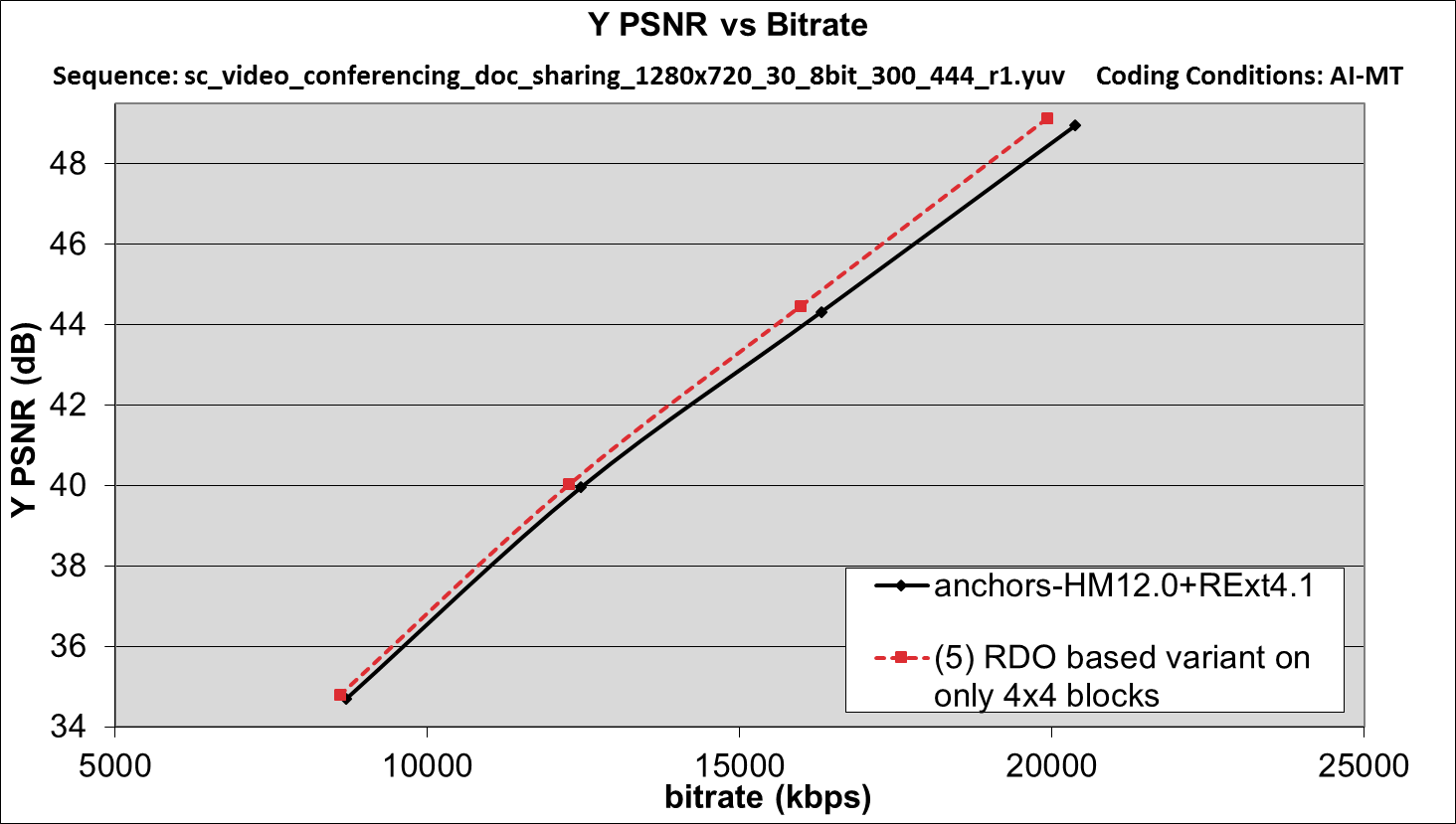}%
			\label{fig:rdc_vc}}}
	\caption{Rate Distortion Curve of ChinaSpeed, sc\_twist\_tunnel and sc\_vc\_doc\_sharing encoded with method \ref{rdo}.}
	\label{fig:rdc}
\end{figure*}

To illustrate the R-D performance under different bitrates, we also plot the R-D curve for three sequences in Fig.~\ref{fig:rdc}.

\subsection{Lossless Coding Results}
\label{LosslessResults}
We apply SAD based method, threshold based method and RDO search method with the lossless coding setting. The results are shown in Table~\ref{tab:results_lossless}. The compression ratio in lossless setting is calculated based on the total bits (including both luma and chroma) of the compressed video and uncompressed video. In this table, the positive number means compression gain expressed as follow:

\begin{equation}
\text{Compression Gain} = \frac{CR_{\text{proposed}} - CR_{\text{anchor}}}{CR_{\text{anchor}}}
\label{eq:losslessgain}
\end{equation}
where $CR_{\text{proposed}}$ and $CR_{\text{anchor}}$ are the compression ratios of our proposed method and the anchor.

\begin{table*}[!t]
\small
	\caption{BD-Bitrate Savings in Lossless Testing. Note That Negative Number Means Compression Gain.}%
	\label{tab:results_lossless}%
	\centering
	\begin{tabular}{|c|c|c|c|c|c|c|c|c|c|c|} \hline
		\multirow{2}{*}{Category} & \multirow{2}{*}{Sequences}
		& \multicolumn{3}{c|}{\tabincell{c}{\ref{sad} SAD \\ variant on \\ only 4$\times$4 blocks}}
		& \multicolumn{3}{c|}{\tabincell{c}{\ref{diff} Pixel difference \\ variant on \\ only 4$\times$4 blocks}}
		& \multicolumn{3}{c|}{\tabincell{c}{\ref{rdo} R-D based \\ variant on \\ only 4$\times$4 blocks}} \\\hhline{|~|~|-|-|-|-|-|-|-|-|-|}
		&       						& AI    	& RA    	& LB & AI    	& RA    	& LB    	& AI    	& RA    	& LB		\\\hhline{|===========|}
		\multirow{11}{*}{\tabincell{c}{Screen \\ Content \\ (All 4:4:4 \\ format)}}
		& ChinaSpeed 				&-0.2&0.0&0.0 & -0.3    &  0.0   	&  0.0 		& -0.4  	& -0.1 	  & 0.0 	\\\hhline{|~|-|-|-|-|-|-|-|-|-|-|}
		& SlideEditing 			&-0.1&0.0&0.0 & -0.1	  &  -0.1   &  0.0  	& -0.4  	& -0.2    & -0.1  \\\hhline{|~|-|-|-|-|-|-|-|-|-|-|}
		& sc\_cad\_waveform &-0.2&0.0&4.4 & -0.3  	&  -0.1   &  4.2    & -3.1  	& -2.6    & -3.2  \\\hhline{|~|-|-|-|-|-|-|-|-|-|-|}
		& sc\_pcb\_layout 	&-0.3&-0.3&-0.1 & -0.6  	&  -0.5   & -0.6    & -7.9 		& -7.3 	  & -6.4  \\\hhline{|~|-|-|-|-|-|-|-|-|-|-|}
		& sc\_ppt\_doc\_xls &-0.6&-0.6&-0.6 & -0.8  	&  -0.8   & -0.8    & -3.1    & -3.1 	  & -2.9  \\\hhline{|~|-|-|-|-|-|-|-|-|-|-|}
		& sc\_programming 	&0.0&0.0&0.0 & -0.1  	&   0.0   &  0.0    & -0.5  	& -0.1 	  & -0.1  \\\hhline{|~|-|-|-|-|-|-|-|-|-|-|}
		& sc\_SlideShow 		&0.0&0.0&0.0 &  0.0 		&   0.0   &  0.0    & -0.1  	& -0.2 	  & -0.1  \\\hhline{|~|-|-|-|-|-|-|-|-|-|-|}
		& sc\_web\_browing 	&0.0&0.0&0.1 &  0.0  	&  -0.1   &  0.0    & -0.7  	& -0.6 		& -0.4  \\\hhline{|~|-|-|-|-|-|-|-|-|-|-|}
		& sc\_wordEditing 	&-0.3&-0.2&-0.1 & -0.1  	&   0.1   & -0.1    & -0.8  	& -0.5 		& -0.4  \\\hhline{|~|-|-|-|-|-|-|-|-|-|-|}
		& sc\_twist\_tunnel &-0.6&-0.7&-0.6 & -3.2 		&  -3.6   & -3.6    & -8.2 	  & -8.2 		& -8.0  \\\Cline{0.9pt}{2-11}
		& \textbf{Average} 	&\textbf{-0.2}&\textbf{-0.2}&\textbf{0.3} & \textbf{-0.5} & \textbf{-0.5} & \textbf{-0.1}  	& \textbf{-2.5} & \textbf{-2.3} & \textbf{-2.2} \\\hhline{|===========|}
		\multirow{4}{*}{\tabincell{c}{Mixture of \\ Screen and \\Natural \\ Content}}
		& BasketballDrillText (4:2:0) &0.0&0.0&0.0	&  0.0  &  0.0   &  0.0  & 0.0   &  0.0		&  0.0	\\\hhline{|~|-|-|-|-|-|-|-|-|-|-|}
		& sc\_map (4:4:4)						&0.0&0.0&0.0	&  0.0  &  0.0   &  0.0  & -1.0   &  -0.4 	& -0.3 	\\\hhline{|~|-|-|-|-|-|-|-|-|-|-|}
		& sc\_vc\_doc\_sharing (4:4:4) &-0.3&-0.1&0.0	& -0.4  & -0.2   & -0.1  & -1.9  &  -1.0 	& -0.5	\\\Cline{0.9pt}{2-11}
		& \textbf{Average} 		 &\textbf{-0.1}&\textbf{0.0}&\textbf{0.0}	& \textbf{-0.1} & \textbf{-0.1} & \textbf{0.0} & \textbf{-1.0} & \textbf{-0.5} & \textbf{-0.3}\\\hhline{|===========|}
		\multirow{6}{*}{\tabincell{c}{Natural \\Content}}
		& Kimono (4:4:4)					&0.0&0.0&0.0		& 0.0     	&  0.0    &	0.0		& 0.0     & 0.0		& 0.0	\\\hhline{|~|-|-|-|-|-|-|-|-|-|-|}
		& ParkScene (4:2:0)			&0.0&0.0&0.0		& 0.0     	&  0.0    &	0.0		& 0.0     & 0.0		& 0.0	\\\hhline{|~|-|-|-|-|-|-|-|-|-|-|}
		& EBUHorse (4:2:2)				&0.0&0.0&0.0		& 0.0   		&  0.0    & 0.0		& 0.0     & 0.0	  & 0.0	\\\hhline{|~|-|-|-|-|-|-|-|-|-|-|}
		& EBUWaterRocksClose (4:2:2) &0.0&0.0&0.0	& 0.0     	&  0.0    & 0.0		& 0.0     &  0.0 	& 0.0	\\\hhline{|~|-|-|-|-|-|-|-|-|-|-|}
		& EBURainFruits (4:2:2)		&0.0&0.0&0.0	& 0.0    		&  0.0    &	0.0 	& 0.0     &  0.0	& 0.0	\\\Cline{0.9pt}{2-11}
		& \textbf{Average} 	& \textbf{0.0} & \textbf{0.0} & \textbf{0.0}	& \textbf{0.0} & \textbf{0.0} & \textbf{0.0} & \textbf{0.0} & \textbf{0.0}	& \textbf{0.0}\\ \hhline{|===========|}
		\multirow{2}{*}{Summary} 	& Enc. Time (\%) & 102    &  100     & 101 & 101    &  101     & 102    &   105    & 101   & 101 	\\
		& Dec. Time (\%) & 100    &  101     & 100 & 101    &  101     & 101    &   101    & 100   & 100 	\\\hline
	\end{tabular}
\end{table*}%

Similar to the lossy coding results, we observe the same trend in lossless coding results. Specifically, the R-D based scheme provides the best compression gains amongst the three proposed schemes, as the correct R-D decisions are made. The Rate-Distortion based scheme performs well on all sequences, and coding gains of 2.5\% (AI),  2.3\%(RA), and 2.2\% (LB) on average are obtained for screen content video. Further, there is no loss on camera-captured content.

\begin{table}[!t]
\footnotesize
	\caption{BD-Bitrate Savings (\%) of each Color Components on YUV 4:4:4 Screen Content Video Sequences. All-Intra, Lossy Coding with Method (5). Negative Number Means Compression Gain.}%
	\label{tab:results_CbCr}%
	\centering
	\begin{tabular}{|c|c|c|c|} \hline
		Sequence                & Y             & Cb            & Cr  \\\hhline{|====|}
		ChinaSpeed 				&-2.1           &-1.4           &-1.2 \\\hhline{|-|-|-|-|}
		SlideEditing 			&-1.1           &-1.0           &-1.1 \\\hhline{|-|-|-|-|}
		sc\_cad\_waveform       &-3.1           &-2.7           &-2.6 \\\hhline{|-|-|-|-|}
		sc\_pcb\_layout 	    &-7.4           &-6.7           &-6.9 \\\hhline{|-|-|-|-|}
		sc\_ppt\_doc\_xls       &-3.4           &-3.0           &-3.0 \\\hhline{|-|-|-|-|}
		sc\_programming 	    &-1.2           &-0.6           &-0.6 \\\hhline{|-|-|-|-|}
		sc\_SlideShow 		    &-0.6           &-0.6           &-0.7 \\\hhline{|-|-|-|-|}
		sc\_web\_browing 	    &-1.4           &-0.8           &-0.9 \\\hhline{|-|-|-|-|}
		sc\_wordEditing 	    &-1.5           &-0.7           &-0.9 \\\hhline{|-|-|-|-|}
		sc\_twist\_tunnel       &-11.5          &-10.2          &-9.9 \\\hhline{|====|}
		\textbf{Average}        &\textbf{-3.3}  & \textbf{-2.8}   &\textbf{-2.8} \\\hhline{|-|-|-|-|}
	\end{tabular}
\end{table}%

\subsection{Gain on Cb and Cr components}
\label{sec:CbCr}
As shown in Table \ref{tab:results_lossy}, all screen content test sequences are YUV 4:4:4 formats. So the performance of the selective interpolation on Cb and Cr components are also important. We show the BD-Bitrate saving of each component in Table \ref{tab:results_CbCr}. Note that in this table, we only show the results of All-Intra Coding with method (5), the RDO-based variant. From this table we can see, our method can work on the Cb and Cr components well (similar gain -2.8\% is achieved.)

\subsection{Hit Ratio}
\label{sec:hitratio}
\begin{figure*}[!t]
	\centerline{\subfloat[ChinaSpeed]{\includegraphics[width=0.2\textwidth]{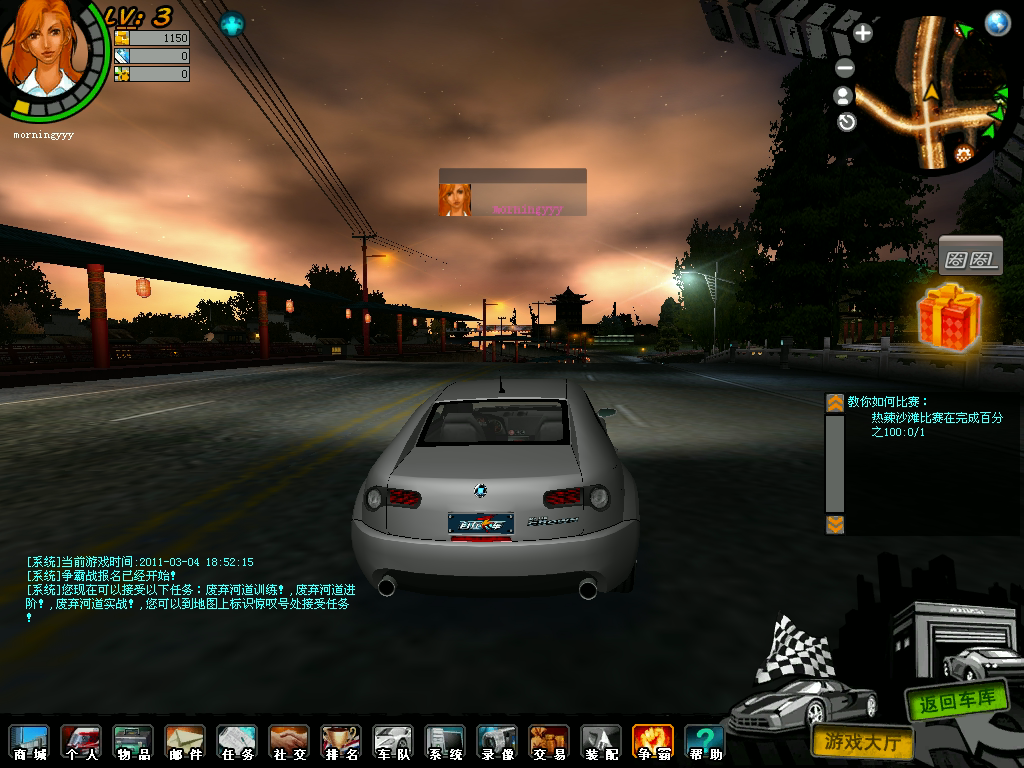}%
			\label{fig:ChinaSpeed}}
		\hfil
		\subfloat[Hit ratio]{\includegraphics[width=0.2\textwidth]{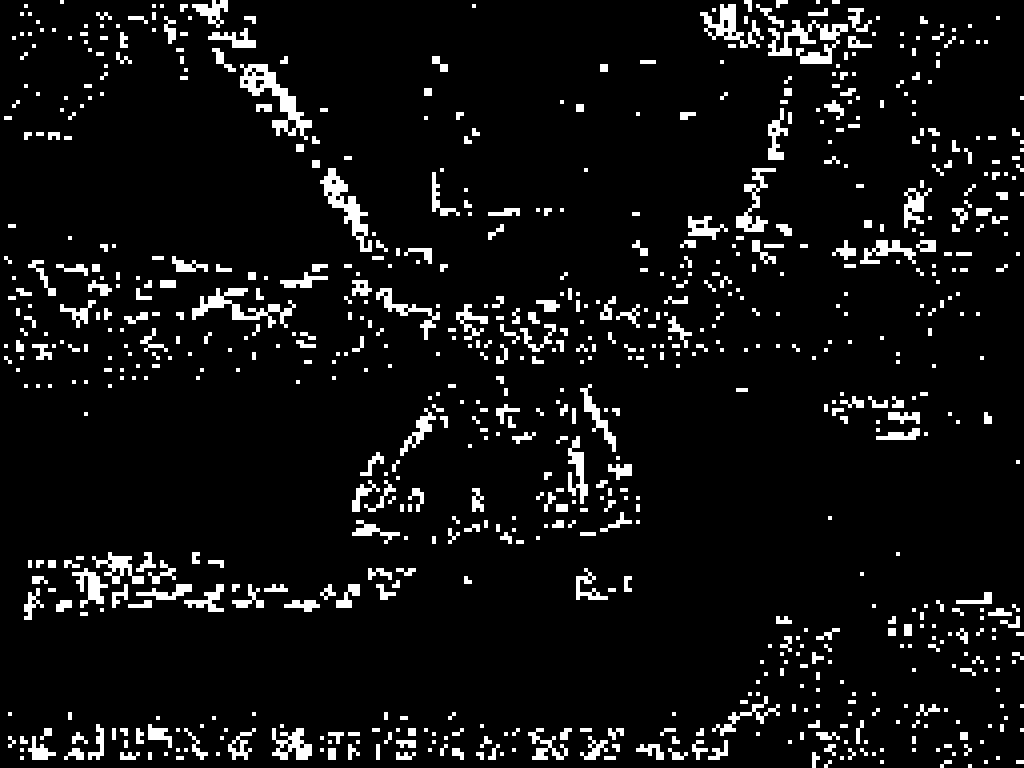}%
			\label{fig:hitratioChiSpeed}}
		\hfil
		\subfloat[sc\_map]{\includegraphics[width=0.24\textwidth]{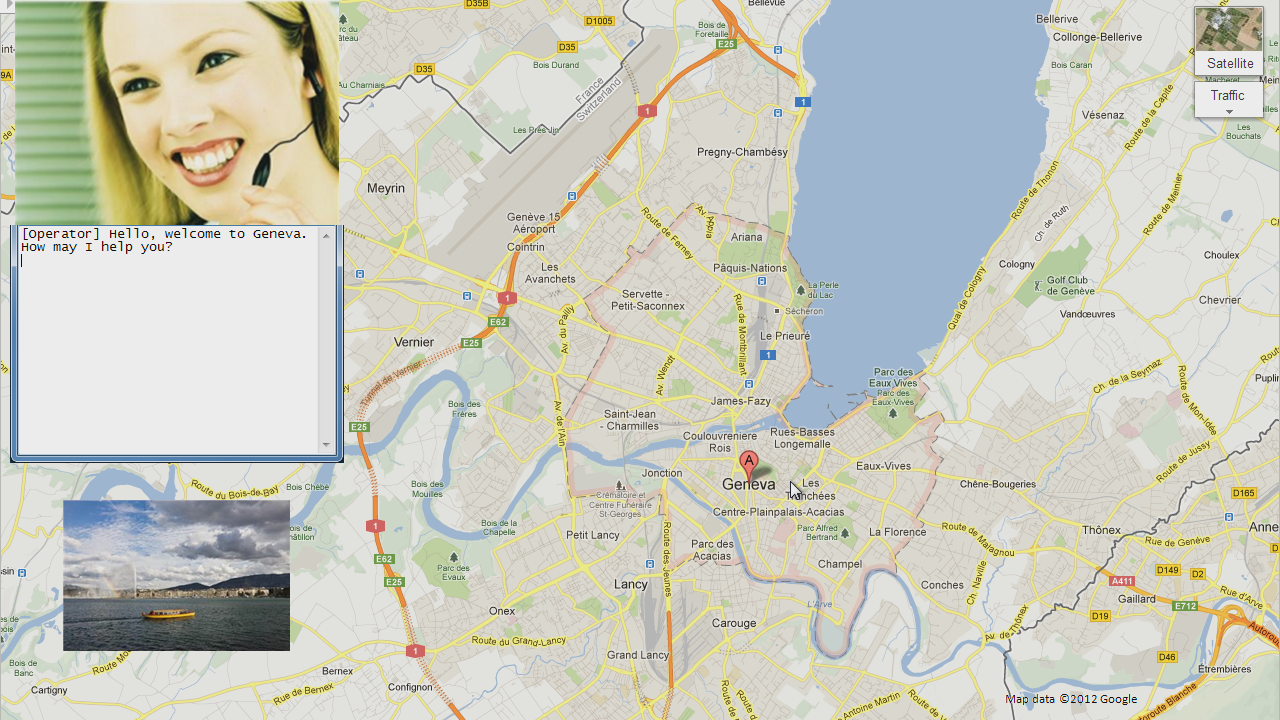}%
			\label{fig:scmap}}
		\hfil
		\subfloat[Hit ratio]{\includegraphics[width=0.24\textwidth]{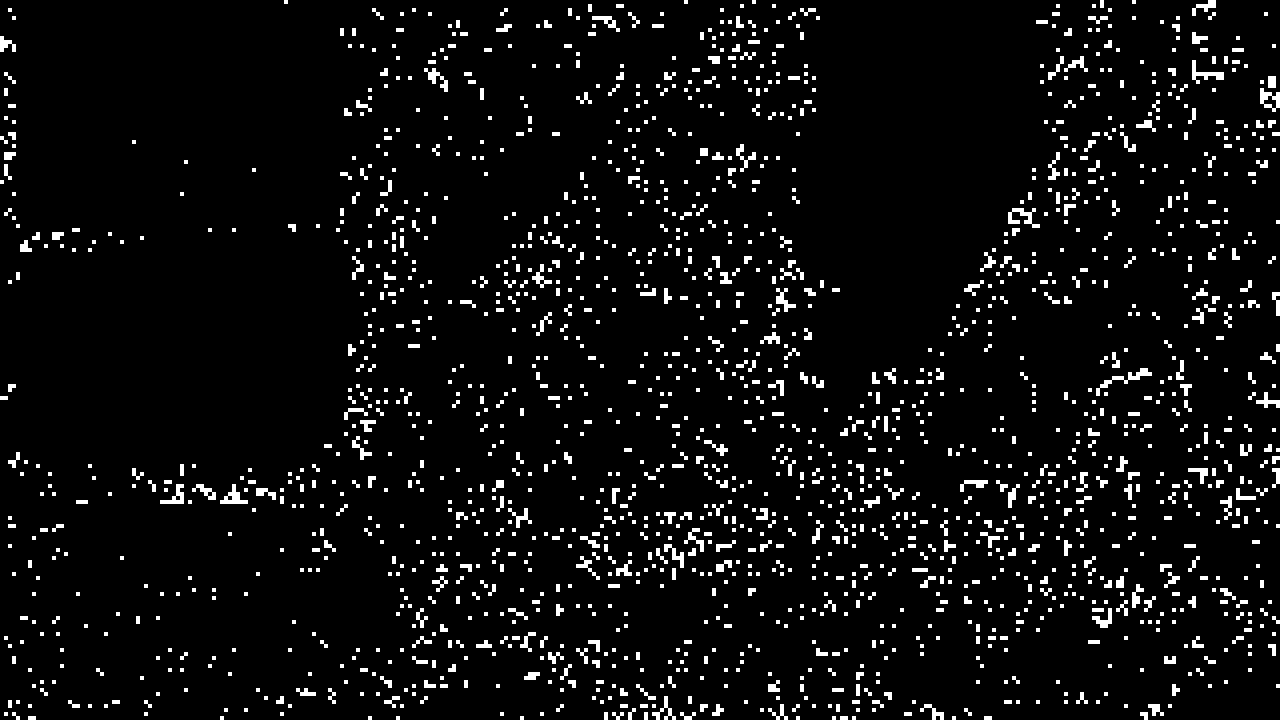}%
			\label{fig:hitratioscmap}}}
	\caption{First frames of two video sequences: ChinaSpeed and sc\_map, and their corresponding hit ratio illustration. The white blocks denote the $4\times4$ blocks using nearest neighbor prediction, and the black blocks use bilinear interpolation, or are IntraBC blocks (Figure best viewed in color.)}
	\label{fig:hitratio}
\end{figure*}

Next, to illustrate that the nearest neighbor prediction scheme is suitable in screen content video, we present the hit ratio information when it is used for two video sequences in
Fig.~\ref{fig:hitratio}. The white blocks correspond to the blocks selected as using nearest neighbor interpolation in the RDO search, and the black for bilinear interpolation, and blocks coded with Intra Block Copy (IntraBC) \cite{joshi2013ahg8} mode. A large portion of the frames, especially the boundary, use the nearest neighbor prediction. Note that in Intra frames, with the advent of IntraBC blocks, the whole block can be predicted from a different block within the frame (similar to motion estimation for Inter frames), and therefore the number of blocks using conventional intra prediction (not the new IntraBC mode) are less. When we disabled IntraBC, the compression gains of nearest neighbor prediction scheme increased further significantly. According to our simulation results, 3.0 \% gain on average for method \ref{rdo} can be further achieved in All Intra Lossy setting when IntraBC is disabled.

\section{Discussion of the Complexity and Parallelism}
\label{sec:complexity}
\subsection{Additional complexity}
\label{sec:add_op}
The additional complexity of the proposed nearest neighbor prediction scheme at the decoder is almost negligible, as the decoder has to only check whether or not to do the nearest neighbor interpolation based on the SAD or the reference pixels, pixel difference, or parse a flag in the R-D variant. It can be seen from the decoding time comparison in Table~\ref{tab:results_lossy} that the increment in decoder run-times is negligible.  At the encoder, the additional complexity of the SAD and pixel difference based variants is again negligible, while the R-D search incurs about 8 \% and 5 \% increase in the run-times for the All Intra lossy and lossless coding settings, respectively due to R-D search for the intra prediction modes. For the RA, and LB settings, the increase in encoder run-times is around 1 \% or less. Compared with some other major SCC tools, e.g., Intra Block Copy \cite{lee2014ahg5} (70\% additional encoding time), Transform Skip \cite{lan2012intra} (14\% additional encoding time) and RDPCM \cite{zhou2011ahg22, joshi2013residual} (4\% addtional encoding time), the additional complexity of our methods (8\% additional encoding time for RDO-based variant and 1\% for other two variants) is acceptable.

\subsection{Parallelism}
\label{sec:para}
Parallelism is used to speed up the whole encoding and decoding process. In hardware implementation, sometimes the original pixels, instead of the reconstructed pixels can be used to do the some mode decisions, so that multiple blocks can be calculated simultaneously. For example, the intra mode can be derived based on the R-D performance of the original reference pixels. In our proposed scheme, RDO-based variant can also be implemented in parallel as the interpolation information is included in the bitstream, but the the SAD and pixel selection variants are not suitable for parallel implementation, since the same decision cannot be derived in the decoder side if the original pixels are used as reference pixels.



\section{Conclusion}
\label{sec:conclusion}
To the best of our knowledge, this is the first paper that 1) points out current HEVC intra prediction scheme with bilinear interpolation does not work efficiently for screen content video and 2) uses different filters adaptively in the HEVC intra prediction interpolation. In this paper, we demonstrate that the NN intra prediction can achieve better coding efficiency when coding screen content video, especially in the presence of sharp edges. We also describe three variants of when to selectively use NN interpolation or bilinear interpolation schemes: SAD based, pixel-difference based, and Rate-Distortion search based. Simulation results show significant gains (3.3\% for All-Intra Lossy coding and 2.5\% for All-Intra Lossless coding) for the proposed nearest neighbor prediction scheme. Future work includes finding more effective implicit criteria to selectively use the NN intra prediction scheme, and using information from already coded neighboring blocks about the interpolation scheme being used in them so as to utilize the correlation during the coding of current block.

\section*{References}

\bibliography{NN_Intra_trans_jour}

\end{document}